\documentclass{article}      
\usepackage{graphicx}
\usepackage{epstopdf}
\title{High $Q^2$ behavior of the electromagnetic form factors in the relativistic hypercentral CQM}

\author{E. Santopinto\\
Istituto Nazionale di Fisica Nucleare, Sezione di Genova, Italy,\\
A. Vassallo, M.M. Giannini\\
Dipartimento di Fisica dell'Universit\`a di Genova\\
and \\
Istituto Nazionale di Fisica Nucleare, Sezione di Genova, Italy, \\
M. De Sanctis\\
Universidad Nacional de Colombia, Bogot\'a (Colombia)
}

\date{}
\begin{document}

\maketitle  

\begin{abstract}
The ratio $R_p$ between the electric and magnetic proton form factors has been recently measured at JLab up to $Q^2=8.5 ~GeV^2$. We have extended the calculation of the nucleon form factors with the hypercentral Constituent Quark Model and compared them with the data on $R_p$ and on the $Q^2$ behavior of the ratio $Q^2 F_2/F_1$. In both cases the theoretical curves agree with the experimental points.
\end{abstract}

The description of the internal nucleon structure has been object of a renewed interest due to the recent polarization experiments on the ratio 
\begin{equation}
R_p~=~\mu_p \frac {G^p_E(Q^2)}{G^p_M(Q^2)}
\end{equation}

\noindent between the electric and magnetic form factors of the proton up to $Q^2 =~ 5.5 ~ GeV^2$ \cite{Milbrath99,Jones:1999,Gayou:2001,Gayou:2002,Pospischil03,Punjabi05}. The measured values of $R_p$ decrease strongly as a function of the momentum transfer $Q^2$, at variance with the widely accepted dipole fit, and there seems to be a tendency toward a dip of the electric form factor, which is expected to occur , if data are extrapolated linearly, at about $Q^2 = 8 ~GeV^2$.

The observed decrease at low $Q^2$ may be ascribed to a relativistic effect. In fact, using the hypercentral Constituent Quark Model (hCQM)~
\cite{pl}, we showed explicitly that a decreasing ratio is obtained \cite{ts99,rap,fra99} if the initial and final nucleon states are boosted to the Breit Frame and
relativistic corrections to the non relativistic current are considered \cite{mds}. The calculation \cite{rap} made use of the nucleon form
factors previously determined in \cite{mds}. 

There are various relativistic calculations based on models for the internal nucleon structure and providing a good description of data, both at low and medium $Q^2$, in some cases presenting also a zero of the electric form factor.

The Skyrme soliton model, with vector meson corrections and with the boosting of 
the initial and final nucleon states  to the Breit Frame, has predicted a decrease of $R_p$ and a zero at  $Q^2=10~GeV^2$ (Holzwarth in Ref.\cite{Holz96}). The dip is due to the fact that the intrinsic Skyrme model form factor  has a zero, which is pushed forward by the boosts; in later versions of the calculation the zero occurs at $Q^2 \approx 16~GeV^2$ \cite{Holz99}.
Using a relativistic light cone constituent quark model Frank {\em et al.} \cite{Frank:1995pv} have calculated the electric and magnetic form factors of the proton. The plot of  the values of $R_p$ derived from their results exhibits  a strong decrease with $Q^2$ which is due to a zero in the
electric form factor at $Q^2~=~6~GeV^2$ \cite{Miller:2002}.
Extracting  $G^p_M$ from the matrix elements of the y-component of the current within a light cone
constituent quark model, a decrease of $R_p$, due to the Melosh rotations, is obtained also by Cardarelli and Simula \cite{Cardarelli:2000}; the decrease is however steeper than the data. A formulation of the internal hadron dynamics based on a Bethe-Salpeter approach and the use of a Vector Meson Dominance (VMD) model for the photon-quark vertex has allowed the Rome group to describe both the pion \cite{demelo_pi} and the nucleon \cite{demelo_N} form factors. In the latter case they obtain a ratio $R_p$ which decrease linearly with $Q^2$, in agreement with data, and passes through zero at about $9 ~GeV^2$.   

The Pavia-Graz group \cite{Wagenbrunn:2000es,Boffi:2001zb} has relativized the chiral CQM \cite{olof} within a point form approach; the resulting values for the ratio $R_p$  deviates from $1$ but as a function of   $Q^2$ the theoretical curve decreases too slowly with respect to the experimental data. The  hypercentral Constituent Quark Model, relativized in a point form approach \cite{ff_07}, leads to a noticeable decrease of $R_p$, but in order to reproduce all the details of data, intrinsic quark form factors had to be introduced \cite{ff_07}. At higher $Q^2$, the decrease is however somewhat softened with respect to the experimental trend.

Besides the calculations based on some microscopic model, there are also various fits performed in a VMD framework. In this respect, it should be quoted the work by Iachello  {\em et al.} \cite{IJL}, who obtained a good reproduction of all the existing nucleon form factor data introducing an
intrinsic form factor to describe the internal structure of the nucleon. If one extrapolates the fitted values at high $Q^2$, the resulting $R_p$ decreases with  $Q^2$ and crosses zero at $Q^2 =8~GeV^2$. This analysis has been extended to higher $Q^2$ taking into account the new data \cite{BI}; the results is a good description of $R_p$, but the zero, if any, is pushed toward values much higher than $10~GeV^2$. The fit by Kelly \cite{Kelly} describes very well the data up to $Q^2~=~5.5~GeV^2$ and presents a zero at about $Q^2 \approx 10~GeV^2$. Also Lomon \cite{Lomon} is able to obtain a good fit and, extrapolating his formulae al higher $Q^2$, one observes a zero between  $13$ and $14~GeV^2$. In Ref. (\cite{Alberico})  Alberico {\em et al.} have performed two fits. In Fit I, the electric form factor $G^p_E(Q^2)$ is forced to depend linearly on $G^p_M(Q^2)$ and therefore $R_p$ crosses zero at $Q^2~=~8~GeV^2$; in Fit II, such linear dependence is relaxed and the data are again very well described up to $Q^2~=~3~GeV^2$, but for higher values of $Q^2$, the decrease of $R_p$ seems to be softened.

It is quite understandable that such a situation enhances the expectation of new data at higher $Q^2$, as planned for instance at Jefferson Lab. Very recently \cite{Puckett10} the polarization measurements have been extended up to $Q^2~=~8.5~GeV^2$ (see full points in Fig. 1). Unfortunately the error bars, specially for the last points, are quite large and there are probably consistency problems with the previous data; nevertheless their average trend seems to indicate a softening of the decrease. If they are confirmed, a zero at  $Q^2~=~8~GeV^2$ is difficult to be expected. It is therefore interesting to compare these new data with the theoretical curves provided by the hypercentral Constituent Quark Model of Ref. \cite{ff_07}.

\begin{figure}[h]

\includegraphics[width=3.in]{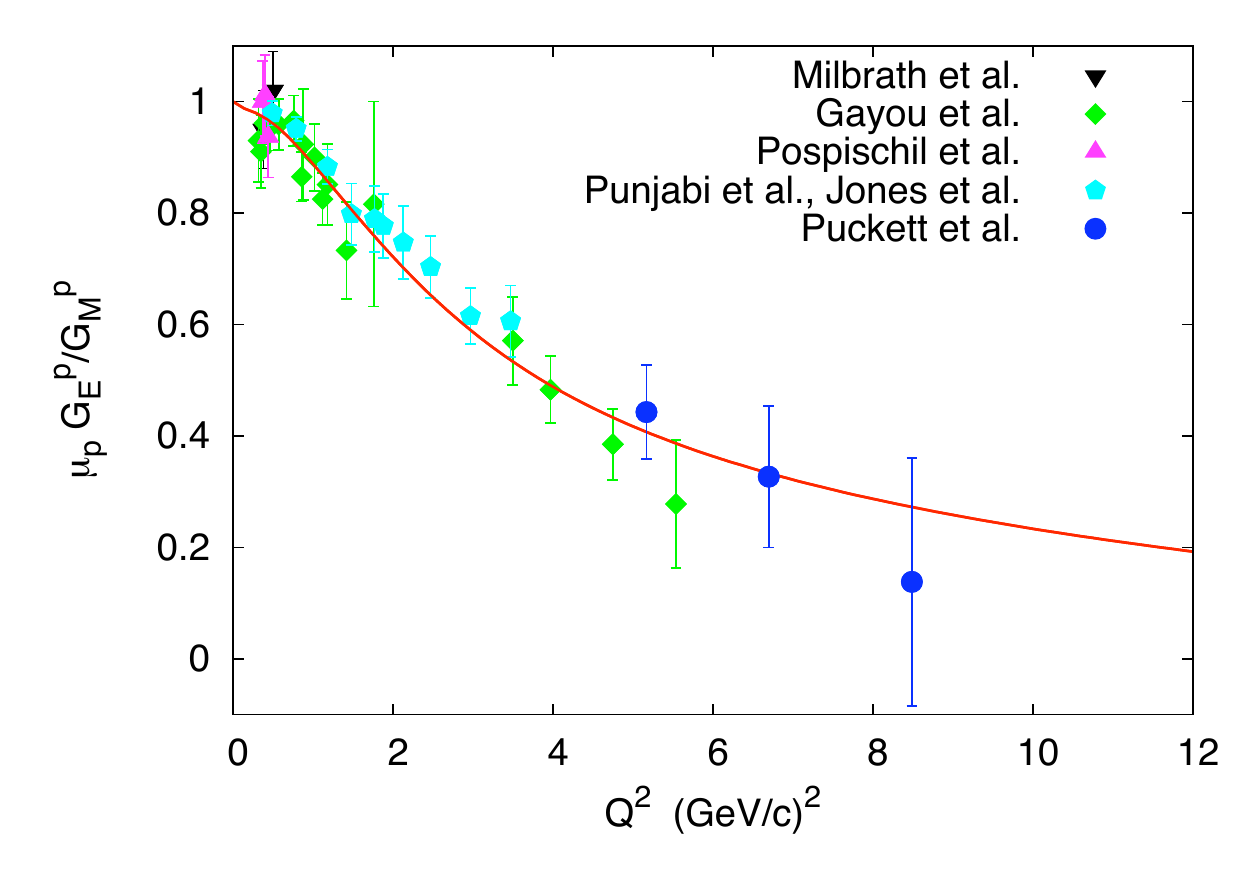}
\caption{(Color on line) The ratio $R_p$ at high $Q^2$ given by extending the calculation performed in \cite{ff_07}. Data are taken from \cite{Milbrath99,Jones:1999,Gayou:2001,Gayou:2002,Pospischil03,Punjabi05,Puckett10} }
\label{R_p}
\end{figure}

In Fig.\ref{R_p}, we report the results of Ref. \cite{ff_07} simply extending the calculations up to $Q^2~=~12~GeV^2$, without any further parameter fitting. The theoretical curve shows that $R_p$ actually continues to decrease but there seems to be no indication of a crossing at finite values of $Q^2$. The agreement with the new data is remarkable.

\begin{figure}[h]
\includegraphics[width=3.in]{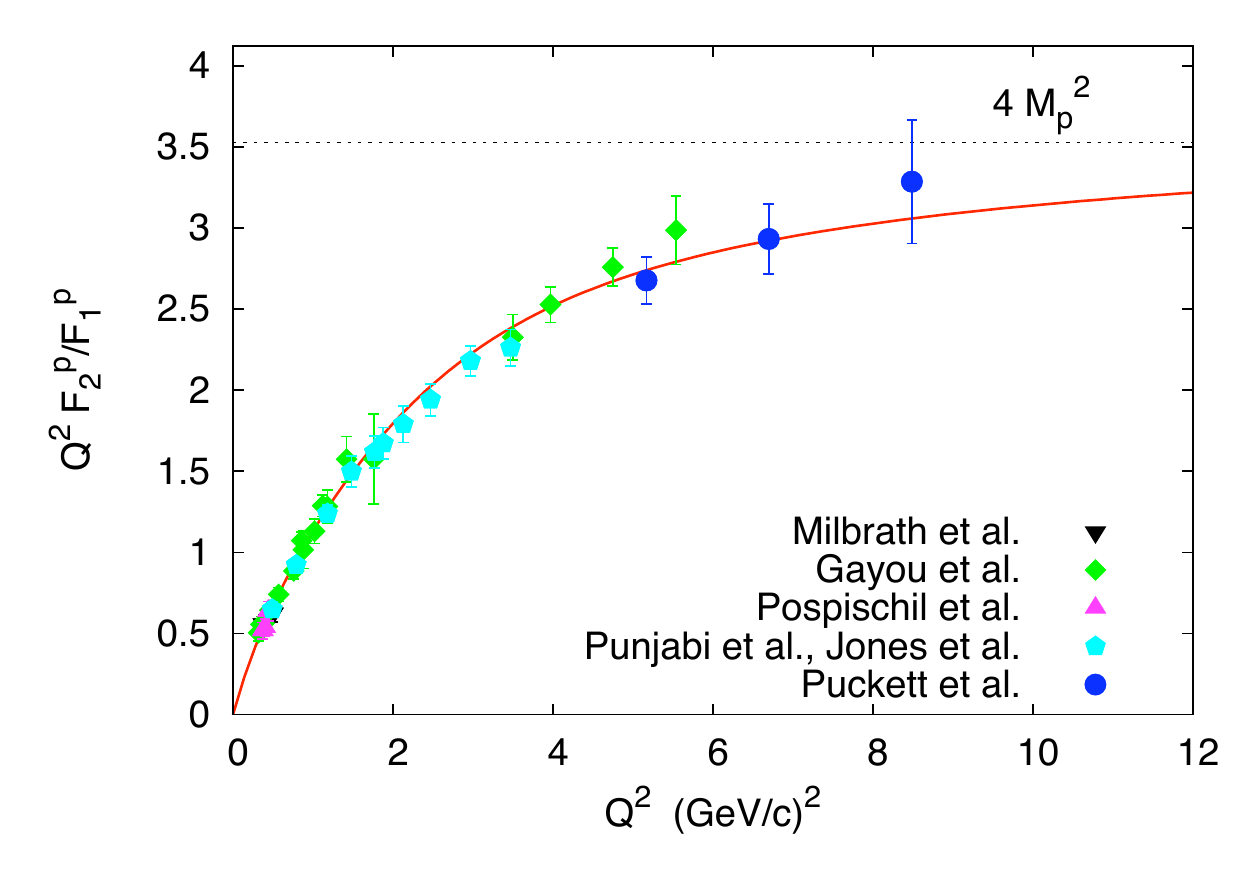}
\caption{(Color on line) The ratio $F_p$ at high $Q^2$ calculated using the theoretical form factors of Ref. \cite{ff_07}. The experimental references are the same as in Fig. \ref{R_p} }
\label{F_p}
\end{figure}

\begin{figure}[h]
\includegraphics[width=3.in]{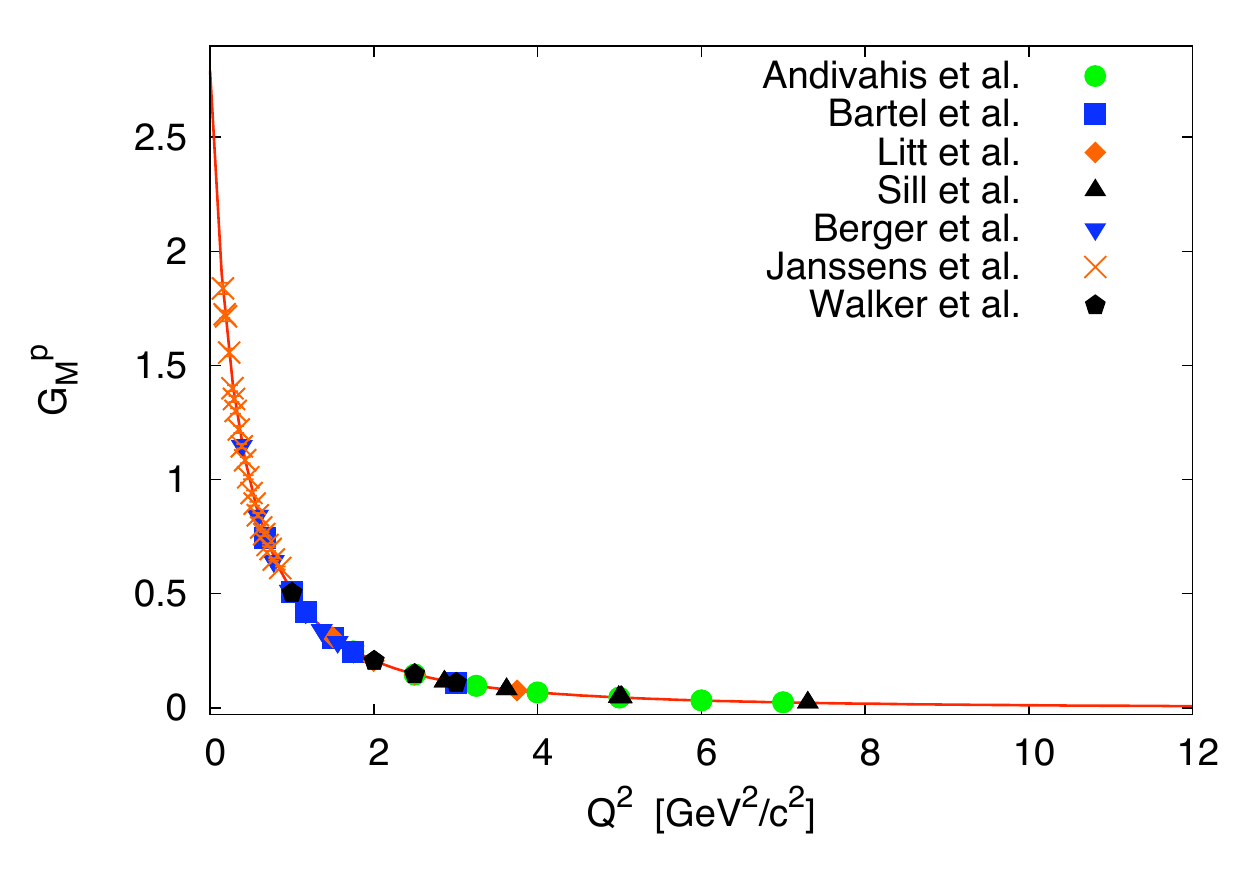}
\caption{(Color on line) The proton magnetic ($G_M^p$) form factor calculated in Ref. \cite{ff_07}.The data  are taken from the reanalysis made in \cite{Brash:2004} of the data from \cite{data}.  }
\label{Gm}
\end{figure}

\begin{figure}[h]
\includegraphics[width=3.in]{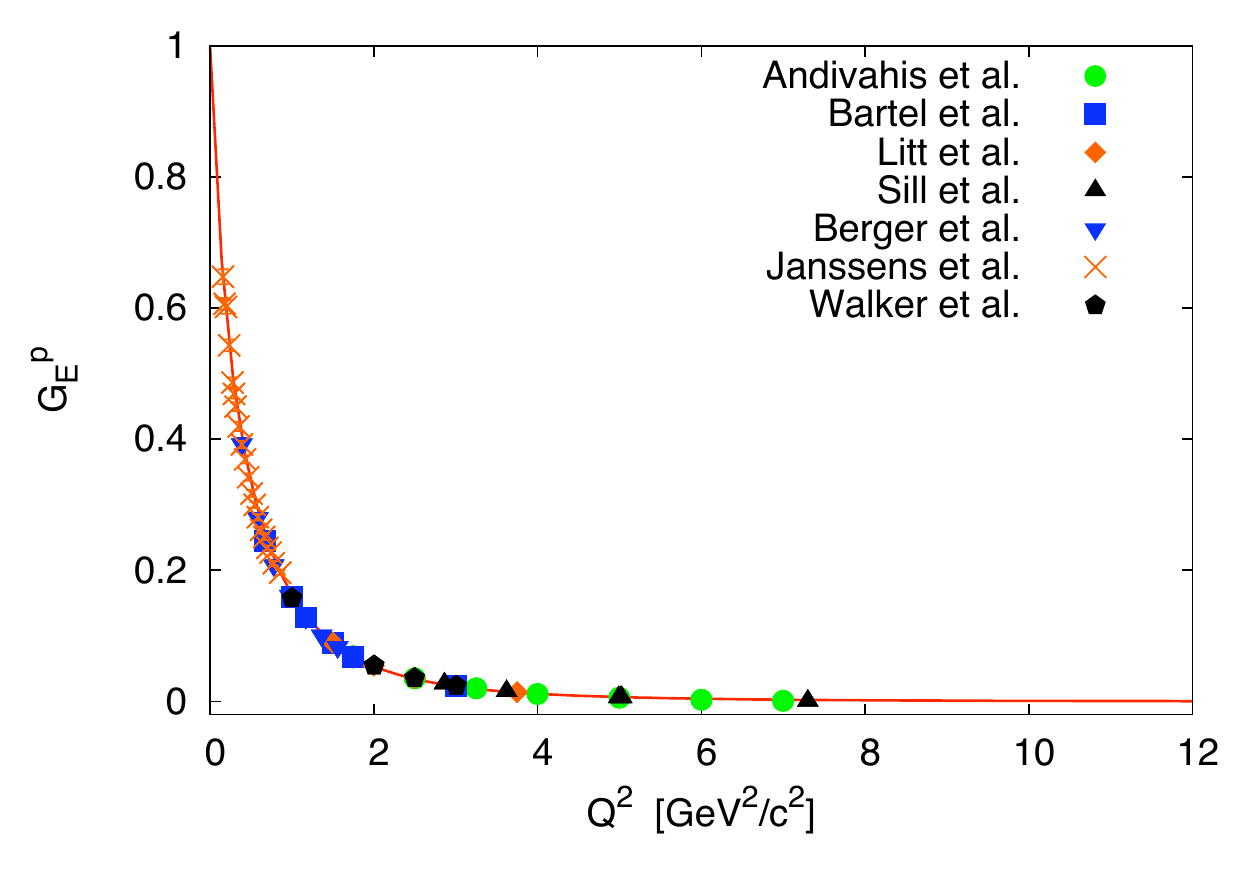}
\caption{(Color on line) The proton electric ($G_E^p$)  form factor calculated in Ref. \cite{ff_07}.The data 
are obtained from the $G_M^p$ data of Fig. (\ref{Gm}) and the fit \cite{Brash:2004} of the JLab data on $R_p$.  }
\label{Ge}
\end{figure}

An alternative way of studying the high $Q^2$ behavior of the proton form factors is that to considering the ratio
\begin{equation}
F_p~=~ Q^2\frac {F^p_2(Q^2)}{F^p_1(Q^2)}
\end{equation}
\noindent It is interesting to note that in correspondence of a zero of the proton form factor $G^p_E(Q^2)$, $F_p$ must cross the value $4 M_p^2$, where $M_p$ is the proton mass. According to the analysis of Ref. \cite{brodsky-farrar,brodsky-lepage}, such ratio should reach an asymptotic constant value.

The ratio $F_p$ calculated using the form factors of Ref.  \cite{ff_07} is reported in Fig. \ref{F_p}. Here again the agreement with data is quite good. The data and the theoretical curve exhibit a trend to saturation, according to the prescription of Ref. \cite{brodsky-farrar,brodsky-lepage}. It is also evident that the existence of a crossing  at $4 M_p^2$, that is the presence of a dip in the electric proton form factor, is still not indicated by the present experimental data,  thus measurements at higher $Q^2$ are necessary.

The fact that the ratios $R_p$ and $F_p$ are reproduced does not ensures that the same may happen for the absolute values $G_E^p$ and $G_EM^p$. For this reason, we report in Figs. (\ref{Gm}) and (\ref{Ge}) the absolute values $G_E^p$ and $G_EM^p$ calculated in Ref.  \cite{ff_07}, showing that also the individual form factors of the proton are quite well described.

The good agreement with data shown in Ref. \cite{ff_07} and in the present work is obtained thanks to the introduction of intrinsic quark form factors, which take into account implicitly any dynamical aspect not included in the theory. Also in other works the good agreement is due to the consideration either of quark from factors (e.g.\cite{Cardarelli:2000}) or considering a vector meson dominance mechanism \cite{Holz96,Holz99,demelo_N,IJL,Kelly,Lomon, Alberico}, pointing toward the necessity of going beyond the three-valence quark approach. A strong indication in this sense is also given by the  application of the hypercentral Constituent Quark Model to the description of the $Q^2$ dependence of the helicity amplitudes  \cite{aie,aie2}. In fact, the emerging picture is that of small quark core, about $0.5 ~ fm$ large, surrounded by a meson cloud or quark-antiquark pairs. As long as these non valence contributions remain in the outer region, they are important at low $Q^2$ \cite{Tiator}. However, the quark-antiquark pair creation mechanism should be introduced in a fully consistent way by unquenching the CQM as in Ref. \cite{BS} and it is expected to be effective also at short distance. Another consequence of the unquenching of the CQM is that baryon states acquire higher Fock components, which certainly contribute to the electromagnetic properties of hadrons and, as noted in \cite{WP12}, may have a quite different high $Q^2$ behavior with respect to the standard three-quark configurations. In this way, the interference between the valence and sea contributions become a possible candidate for the generation of a dip in the electric form factors.

All these aspects will be hopefully clarified by the future experiments planned al Jefferson Lab with the $12 ~ GeV$ upgraded accelerator.

\end{document}